\documentclass[conference]{IEEEtran}
\IEEEoverridecommandlockouts
\usepackage{etex}
\reserveinserts{100}
\usepackage{mathtools, graphics, theorem, times, amsfonts, graphicx, amssymb, cite}
\usepackage[outdir=./]{epstopdf}
\usepackage{tikz}
\usetikzlibrary{shapes,arrows}
\usetikzlibrary{matrix} 
\usetikzlibrary{decorations.markings} 
\usetikzlibrary{calc} 
\usetikzlibrary{decorations.pathmorphing,snakes, positioning, decorations.pathreplacing}

\usetikzlibrary{fit}
\usepackage{pgfplots}
\usepackage{color}
\usepackage{setspace}
\usepackage{hyperref}
\usepackage{multirow}
\usepackage{rotating}
\usepackage{comment}
\usepackage{caption}
\usepackage{subcaption}
\usepackage[affil-it]{authblk}
\usepackage{bm}
\usepackage{algorithm}
\usepackage{algorithmic}
\usepackage{enumerate}
\usepackage{morefloats}
\usepackage{setspace}

\input{mysymbol.sty}
\usepackage{needspace}

\def\E{\mathbb{E}}

\begin{document}

\title{Communication-Control Co-design in  \\ Wireless Edge Industrial Systems}
%

\author{Mark Eisen \quad Santosh Shukla \quad Dave Cavalcanti \quad Amit S. Baxi
\thanks{M. Eisen and D. Cavalcanti are with Intel Labs, Hillsboro, OR, USA and S. Shukla and A. S. Baxi are with Intel Labs, Bangalore, India. Email: \{mark.eisen, santosh.shukla, amit.s.baxi, dave.cavalcanti\}@intel.com }}


\thispagestyle{empty}
\maketitle

\begin{abstract}
We consider the problem of controlling a series of industrial systems, such as industrial robotics, in a factory environment over a shared wireless channel leveraging edge computing capabilities. The wireless control system model supports the offloading of computational intensive functions, such as perception workloads, to an edge server. However, wireless communications is prone to packet loss and latency and can lead to instability or task failure if the link is not kept sufficiently reliable. Because maintaining high reliability and low latency at all times prohibits scalability due to resource limitations, we propose a communication-control co-design paradigm that varies the network quality of service (QoS) and resulting control actions to the dynamic needs of each plant. We further propose a modular learning framework to solve the complex learning task without knowledge of plant or communication models in a series of learning steps and demonstrate its effectiveness in learning resource-efficient co-design policies in a robotic conveyor belt task.
\end{abstract}


\section{Introduction}
Edge computing has recently become a powerful tool in the flexible virtualization and offloading of complex and computationally heavy tasks used in industrial and robotic control scenarios  \cite{dey2016robotic, rahman2018communication}. Such advancements follow similar trends in smart factories and Industry 4.0, which envision large scale networked control systems that offload control, perception, or analytics functions to the edge over a wireless network. However, this introduces new significant challenges in the operation of complex industrial control systems. Namely, wireless edge control requires the design of controllers to operate under delays and packet loss, as well as the design of a resource management system to share the wireless network resources used to access the edge.

There is then considerable interest in determining how to maintain strong wireless network and control task performance in these settings. Many radio resource allocation schemes in the form of wireless scheduling techniques have been proposed to provide reliability, or quality of service (QoS), to users across the network in the form of throughput, fairness and/or latency~\cite{cao2001scheduling,ergen2003qos,liu2003delay, yaacoub2012survey}. These methods, however, ignore the underlying needs of the control applications they support. Industrial IoT systems are modeled most often as a series of dependent or independent control systems, where each plant represents a device, e.g. robotic agent. Control-aware schedulers have been developed that provide access to the communication medium dynamically at each step based on current control system states \cite{ mamduhi2014event,shi2011optimal,han2017optimal, GatsisEtal15, eisen2019control}. In emerging applications such as robotics, plant dynamics and performance targets are not so easily modeled, which has further motivated the use machine learning and reinforcement learning techniques to derive resource allocation policies \cite{demirel2018deepcas,leong2020deep,eisen2020network}. 

The methods described above are each limited in their focus on resource allocation and scheduling methods that adapt to control system state. A more complete formulation of the problem addresses the \emph{co-design}, or joint optimization, of network resource allocation and control system operation. The communication-control co-design problem is challenging but has been addressed with heuristics for cases with simple models \cite{molin2009lqg, gatsis2014optimal}, with limited applicability in complex industrial control systems such as robotics \cite{mechraoui2009co,hu2019co}. Due to the limitations in modeling complex systems, the co-design problem has subsequently been tackled using reinforcement leaning techniques \cite{baumann2018deep,lima2020model}.

Existing co-design formulations are limited from their application to realistic wireless protocols, such as 5G and WiFi 6, which use complex OFDMA based scheduling architectures that are not easily addressed with machine learning. In this paper we propose a new formulation of the co-design problem that combines QoS-aware control with dynamic QoS adaptation, in which high-level network performance targets such as packet loss probability and latency are determined for each plant based on their current state. We begin by formulating the dynamics of a control system that is impacted by potential packet loss and delayed packets (Section \ref{sec_problem}). We define the optimal co-design policy in a manner that maximizes network efficiency while meeting control system task requirements (Section \ref{sec_pf}). The resulting constrained Markov decision process cannot be solved without explicit modeling of the plant dynamics and is moreover challenging to solve for complex systems such as robotics, and as such we leverage a model-free reinforcement learning methodology to optimize co-design policies (Section \ref{sec_drl}). We further detail how dynamic and heterogeneous network QoS requirements can be considered in high level scheduling heuristics (Section \ref{sec_scheduling}). The performance of the policy is studied for a use-case of particular interest in industrial settings, namely an industrial robot and conveyor belt system, and superior performance of the learning co-design solution is established for this practical setting (Section \ref{sec_num}), including performance analysis in a simulated WiFi 6 network (Section \ref{sec_wifi}). 

\section{System Model}\label{sec_problem}

\begin{figure}
\centering
 \pgfdeclarelayer{bg0}    
\pgfdeclarelayer{bg1}    
\pgfsetlayers{bg0,bg1,main}  

\tikzstyle{block} = [draw,rectangle,thick,
text height=0.2cm, text width=0.7cm, 
fill=blue!30, outer sep=0pt, inner sep=0pt]
\tikzstyle{dots} = [font = \large, minimum width=2pt]
\tikzstyle{dash_block} = [draw,rectangle,dashed,minimum height=1cm,minimum width=1cm]
\tikzstyle{smallblock} = [draw,rectangle,minimum height=0.5cm,minimum width=0.5cm,fill= green!30, font =  \scriptsize]
\tikzstyle{smallcircle} = [draw,ellipse,minimum height=0.1cm,minimum width=0.3cm,fill= yellow!40, font =  \scriptsize ]
\tikzstyle{connector} = [->]
\tikzstyle{dash_connector} = [->,thick,decorate,decoration={snake, amplitude =1pt, segment length=8pt}, magenta]
\tikzstyle{branch} = [circle,inner sep=0pt,minimum size=1mm,fill=black,draw=black]

\tikzstyle{vecArrow} = [thick, decoration={markings,mark=at position
   1 with {\arrow[semithick]{open triangle 60}}},
   double distance=1.4pt, shorten >= 5.5pt,
   preaction = {decorate},
   postaction = {draw,line width=1.4pt, white,shorten >= 4.5pt}]

\begin{tikzpicture}[scale=.5, blocka/.style ={rectangle,text width=.9cm,text height=0.6cm, outer sep=0pt}]
 \small

    \matrix(M)[ampersand replacement=\&, row sep=2.5cm, column sep=4pt] {
    
    \node[smallcircle, align=center] (CS1) {Plant \\ {1}};\&\node[smallblock, align=center] (SS1) {Sensor \\{1}};\&
    \node[smallcircle, align=center] (CS2) {Plant \\{2}};\&\node[smallblock, align=center] (SS2) {Sensor \\{2}};\&\&
    \node(d1) {$\cdots$};\&
    \node[smallcircle, align=center] (CSm) {Plant \\\textit{m}};\&\node[smallblock, align=center] (SSm) {Sensor \\ \textit{m}};
    \\
    \node[blocka] (R1) {};\&\node[blocka] (R12) {};\&
    \node[blocka] (R2) {};\&\node[blocka] (R22) {};\&\&
    \node[blocka] (d3) {};\&
    \node[blocka] (Rm) {};\&\node[blocka] (Rm2) {};
    \\
    };

    \node[block] (outer) [fit=(R1.north west) (d3) (Rm2.south east)] {};
    
    \node[align=center, scale =1] at (outer.center) {Edge Processor};
    
    \draw [->, thick, red] (R1) -- node[right]{$\bby_{1,t}$} (CS1);
    \draw [->, dashed, thick] (SS1) -- node[right]{$\bbx_{1,t}$} (R12);
    \draw [->, thick, red] (R2) -- node[right]{$\bby_{2,t}$} (CS2);
    \draw [->, dashed, thick] (SS2) -- node[right]{$\bbx_{2,t}$} (R22);
    \draw [->, thick, red] (Rm) -- node[right]{$\bby_{m,t}$} (CSm);
    \draw [->, dashed, thick] (SSm) -- node[right]{$\bbx_{m,t}$} (Rm2);
%

%
%
%
%
				
		%
		\begin{pgfonlayer}{bg1}
		\node(shared) [fill=red!10, fit={($(CS1.south) + (-15pt, -10pt)$) 
		($(CS2.south) + (-10pt, -10pt)$)
		($(CSm.south) + (+20pt, -10pt)$)
		($(R1.north) + (-15pt, +10pt)$)
		($(R2.north) + (-10pt, +10pt)$)
		($(Rm2.north) + (+20pt, +10pt)$)
		}] {};
		\end{pgfonlayer}
		
		\node[align=center, red!50](shared_medium) at ($(shared.east) + (-150pt, 0pt)$) {Shared \\ Wireless \\ Medium};

\coordinate (FIRST NE) at (current bounding box.north east);
   \coordinate (FIRST SW) at (current bounding box.south west);

	\pgfresetboundingbox
   \useasboundingbox ($(FIRST SW) + (+0pt,0)$) rectangle (FIRST NE);

\end{tikzpicture}
\caption{Wireless control system. Sensing data is processed into state information $\bbx_{i,t}$ at the edge. States of multiple control systems are sent over a shared wireless channel and received by plants as observations $\bby_{i,t}$, which as then used to determine local control inputs $\bbu_{i,t} = \bbg_i(\bby_{i,t})$. }
\label{fig_wcs}
\end{figure}
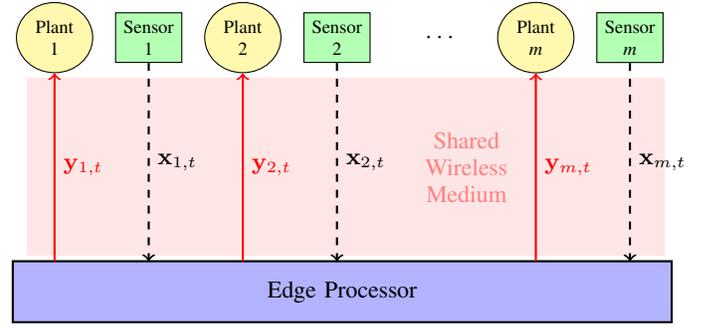

Consider a system of $m$ independent industrial control, or robotic, plants, where each plant $i=1,\hdots,m$ maintains a state variable $\bbx_{i} \in \reals^p$. The dynamics evolve over a discrete time index $t$.  Applying an input $\bbu_{i,t} \in \reals^q$ causes the state to evolve based on some non-linear dynamics,
\begin{align}\label{eq_control_orig}
\bbx_{i,t+1} &= \bbf_i ( \bbx_{i,t}, \bbu_{i,t}) + \bbw_{i,t},
\end{align}
where $\bbf_i : \reals^{p} \times \reals^{q} \rightarrow \reals^{p}$ defines the potentially nonlinear dynamics of plant $i$, and $\bbw_{i,t} \in \reals^{p}$ is a random i.i.d. disturbance with mean $\bb0$ and co-variance $\bbW_i$ that captures potential stochastic variations in the dynamics. As illustrated in Figure \ref{fig_wcs}, the state of each plant is captured by a sensor and transmitted to a centralized edge processor over a wireless medium. The processing at the central server is to perform virtualized robotic perception tasks that extract plant state information $\bbx_{i,t}$ from the sensor measurements.

The wireless network used to connect the sensing and actuation operations adds an inherent disturbance to the idealized control loop model. This is due to the fact that resource limitations in a wireless network introduce non-ideal behaviors to the control cycle, namely \emph{packet loss} and \emph{latency}. The former is caused by potential channel degradation in wireless transmissions while the latter is caused by non-negligible delay in medium access and transmission times.

These non-ideal behaviors are captured by the observation signal $\bby_{i,t}$ received by plant $i$ and time $t$. The packet loss for this control loop is given by the binary indicator $\delta_{i,t} \in \{0,1\}$, with 0 indicating a packet loss. Likewise, the latency is captured by a discrete-valued $\tau_{i,t} \in \mathbb{Z}_+$.  At time $t$, the total delay experience, or the ``age'' of the information received by the plant, is denoted by $\zeta_{i,t}$, and given by the timestamp of the most freshly received state information, i.e.,
\begin{equation}\label{eq_time}
\zeta_{i,t} = t - \max_{t' \leq t}\left\{ t' \mid t' +\tau_{i,t'} \leq t, \delta_{i,t'} =1 \right\}.
\end{equation}
We can thus denote the observed state by plant $i$ as
\begin{equation}\label{eq_output}
\bby_{i,t} = \bbx_{i,t-\zeta_{i,t}}.
\end{equation}
Each plant $i$ has access to a local control policy $\bbg_i: \reals^{p} \rightarrow \reals^{q}$, which is subsequently used to compute control inputs based on observations, i.e.
\begin{equation}\label{eq_control}
\bbu_{i,t} = \bbg_i(\bby_{i,t}).
\end{equation}
Observe in \eqref{eq_output} that, under ideal conditions, $\delta_{i,t}=1$ and $\tau_{i,t} = 0$, and thus perfect state information is known at the controller, i.e. $\bby_{i,t} = \bbx_{i,t}$. Packet loss does not occur deterministically but with a given probability $q_{i,t} \in [0,1]$, which in communication terms denotes the \emph{reliability} of the packet transmission. In other words, the packet loss indicator is drawn randomly from a Bernoulli distribution as
\begin{equation}\label{eq_d}
\delta_{i,t} \sim \text{Bernoulli}(q_{i,t}).
\end{equation}
In general, however, both the reliability and latency may vary due to resource availability in the communication network. In wireless networks, physical channel effects known as fading will cause even further random fluctuations in these phenomena. Together, these variations constitute a varying \emph{quality of service} (QoS) of the communication system. The QoS in wireless networks is then inherently random but can be controlled, up to some limits, through the allocation of radio resources at the physical (PHY) and medium access control (MAC) layers.

Clearly, the controller's access to accurate state information via the network QoS has a measurable impact on the performance of the control system, either in terms of its ability to stabilize the plant or complete a given task. With better QoS the observations $\bby_{i,t}$ become closer to the true plant state $\bbx_{i,t}$, which renders the system easier to control. Standard networked control systems are typically designed to maintain a fixed high target reliability $q_{i,t} \approx 1$ and low latency $\tau_{i,t} \approx 0$ with high probability---see, e.g. \cite{weiner2014design} . This allows control policies designed for ideal conditions, which are well-studied and easier to derive, to be applied. However, in wireless networks, fundamental limits exist on the QoS that can be provided under a given set of resources, e.g. system bandwidth, channel variations, and access time. Enforcing high reliability and low latency subsequently places a large resource requirement on the network or, alternatively, limits the amount of control loops that can be supported in a shared channel with limited resources.
 
For wireless control systems to reach the scale of modern industrial IoT or robotics applications, it is advantageous to loosen the tight requirements on network QoS so as to limit the cost of the supporting network infrastructure and resources. In particular, control systems can often tolerate lost or delayed state information, the frequency of which depends upon the particular plant dynamics as well as internal state estimation capabilities. While some recent works attempt to characterize a fixed reliability or latency the plant dynamics can support to maintain desirable performance \cite{GatsisEtal15, chang2019optimizing}, more generically we may consider \emph{state-dependent} QoS requirements that, for instance, may increase as the estimation error increases or the plant grows closer to an undesirable region---see, e.g. \cite{eisen2019control, chang2019dynamic}. From the communications standpoint, this corresponds to MAC-level radio resource allocation that allows for tuning variable QoS requirements at the packet or frame level. 

Likewise, closed loop controllers designed for perfect communication links are likely no longer optimal under the necessarily limited network performance. While ideal closed loop controllers remain optimal under lossy communications in some specific cases, e.g. linear plants \cite{bar1974dual, molin2012optimality}, this is generally not the case.  Thus, control performance can be further enhanced with precise awareness of varying QoS requirements and limitations. Together, the simultaneous design of dynamic ``control-aware'' QoS requirements in the network with the design of ``QoS-aware'' controllers constitute a \emph{co-design problem} for wireless control systems. We proceed in the next section to derive the optimal co-design for a system of generic non-linear plants described in \eqref{eq_control_orig}-\eqref{eq_control}.

\section{QoS-Aware Optimal Co-design}\label{sec_pf}
In this paper we devise a \emph{dynamic} QoS framework in which targeted wireless network utilization is adaptive and dependent upon the current state and dynamics of each plant. As discussed previously, the QoS for the link between the AP and plant $i$ at time $t$ is characterized by a reliability $q_{i,t}$ and latency $\zeta_{i,t}$. As the intention is to allow these targets to vary with the state of the plant, we consider parameterized policies $\pi^q_i(\cdot; \bbtheta^q_{i})$ and $\pi^{\tau}_i(\cdot; \bbtheta^{\tau}_{i})$ that determine the respective reliability and latency needed in the state transmission, i.e., 
\begin{align}
q_{i,t} &:= \pi^q_i(\bby_{i,t-1}, \zeta_{i,t-1}; \bbtheta^q_i), \label{eq_q_pol}, \\
\tau_{i,t} &:= \pi^\tau_i(\bby_{i,t-1}, \zeta_{i,t-1}; \bbtheta^\tau_i), \label{eq_t_pol}.
\end{align}

Observe in \eqref{eq_q_pol}-\eqref{eq_t_pol} that the policies $\pi_i^q$ and $\pi_i^{\tau}$ are functions of the previous measurements of the plant and its total delay, and are assumed to have specific functional forms, e.g. deep neural networks (DNNs), parameterized by vectors $\bbtheta_i^q \in \reals^{n^q_i}$ and $\bbtheta_i^\tau \in \reals^{n_i^\tau}$, respectively. Together, these policy decisions characterize the network-level QoS provided to plant $i$ at time $t$, and subsequently its quality of measurement given by \eqref{eq_output}.

Just as the wireless network can be provided control-aware intelligence to adapt its performance, so too can the plant controller be provided a QoS-awareness to enhance performance or mitigate the effects of lossy measurements. Consider the standard control policy in \eqref{eq_control} which determines control inputs based solely upon the observation $\bby_{i,t}$ but otherwise agnostic to the underlying manner in which it has been impacted by wireless transmission. Using MAC-layer feedback, such as frame acknowledgment and time stamping, the plant may recover the QoS parameters upon packet receipt and augment its local control policy as
\begin{equation}\label{eq_c_pol}
\bbu_{i,t} := \bbpi^c_i(\bby_{i,t}, \zeta_{i,t}; \bbtheta^c_i).
\end{equation}
Observe in \eqref{eq_c_pol} that, in addition to adding QoS parameters $\delta_{i,t}$ and $\tau_{i,t}$ as inputs, we have further modified the control policy in \eqref{eq_control} by restricting to a specific functional form given by $\bbpi^c_i$ and parameterized by the vector $\bbtheta^c_i \in \reals^{n_i^c}$.

With the policy definitions in \eqref{eq_q_pol}-\eqref{eq_c_pol}, we obtain a series of functions that govern the autonomous system behavior of the wireless edge control system network. It remains to determine how to instantiate such policies to achieve optimal system performance. To do so, we formulate an optimal co-design problem that determines the policy functions via the associated policy parameters $\{ \bbtheta_i^q, \bbtheta_i^\tau, \bbtheta_i^c\}_{i=1}^m$. As previously described, increased QoS on any particular link cannot be achieved arbitrarily, but comes at the expense of wireless resource utilization. We thus consider a loss function $C_i:[0,1] \times \reals_+ \rightarrow \reals$ that measures the cost of achieving a set of QoS targets $\{q, \tau\}$ for plant $i$. Further consider a control-system based loss $J_i(\bbx)$ that measures the performance of plant $i$ when in state $\bbx$.

The objective is to find the minimum allowable QoS performance measures, adapting to system state, such that the necessary control-level performance---captured by a maximum loss $J^{\max}_i$---can be guaranteed using an appropriate QoS-aware control policy. The full co-design policy for plant $i$ is specified by the complete parameter vector $\bbtheta_i^* := [\bbtheta_i^{q*}; \bbtheta_i^{\tau*};\bbtheta_i^{c*}] \in \reals^{n_i}$ and can be defined as the solution to a constrained Markov decision process (MDP) over some finite horizon $T$, i.e.
\begin{subequations}
\begin{align}\label{eq_problem}
\bbtheta_i^* := &\argmin_{\bbtheta \in \reals^{n_i}} \E_{\bbpi_{\bbtheta}} \left[ \sum_{t=0}^T C_i \left( \pi_{\bbtheta}(\bby_{i,t})\right) \right], \\
&\st \quad \E_{\bbpi_{\bbtheta}} \left[ \sum_{t=0}^T  J_i \left( \bbx_{i,t} \right)   \right] \leq J_i^{\max}. \label{eq_const}
\end{align}
\end{subequations}
We use the notation $\E_{\bbpi_{\bbtheta}}[\cdot]$ to denote the expected value under a policy specified by parameter vector $\bbtheta = [\bbtheta_i^q; \bbtheta_i^{\tau}; \bbtheta_i^c]$. 

We emphasize that the formulation of the QoS-aware co-design problem in \eqref{eq_problem}-\eqref{eq_const} is distinct from other related formulations in wireless control---e.g. \cite{lima2020model}---in its basis in abstract communication parameters rather than explicit scheduling decisions at the radio resource level. This approach is beneficial in a number of key aspects that aid in finding solutions in practical scenarios, particularly in its (i) scalability and (ii) agnosticism to specific wireless configurations or standards. 

Regarding (i), observe that the co-design problem is formulated in \eqref{eq_problem}-\eqref{eq_const} for a single control plant. Thus, the optimization is inherently invariant to the number of plants in the network, while optimal co-design policies can be reused for all plants of equivalent dynamics. Regarding (ii), the optimization of high level QoS parameters further allows the designed policy to be used across a variety of wireless networks and standards, such as WiFi and cellular. QoS-based outputs moreover have a simple structure that are amendable to action spaces typical in, e.g., neural network policies. The dynamic QoS outputs in \eqref{eq_q_pol} and \eqref{eq_t_pol} can be subsequently used as inputs to QoS-aware scheduling algorithms \cite{ergen2003qos,andrews2001providing, eisen2019control }.

Significant challenges remain in finding solutions to \eqref{eq_problem}-\eqref{eq_const} for robotic systems. The complexity of non-linear robotic system dynamics as well as the often qualitative nature of robotic task performance metrics render direct application of optimization methods often intractable. In the next section, we detail a machine learning-based framework that reconstructs the co-design problem into a small series of subproblems that can solved in a model-free manner. 

\section{Model-Free Learning Framework}\label{sec_drl}

\begin{figure}
\centering
 \usetikzlibrary{shapes,arrows}

\tikzstyle{decision} = [diamond, draw, fill=blue!20, 
    text width=4.5em, text badly centered, node distance=3cm, inner sep=0pt]
\tikzstyle{block} = [rectangle, draw, fill=blue!20, 
    text width=5em, text centered, rounded corners, minimum height=2em]
\tikzstyle{block2} = [rectangle, draw, fill=red!20, 
    text width=6em, text centered, rounded corners, minimum height=8em]
\tikzstyle{line} = [draw, -latex']
\tikzstyle{cloud} = [draw, ellipse,fill=red!20, node distance=3cm,
    minimum height=2em]
    
\begin{tikzpicture}[node distance = 2cm, auto]
    \footnotesize
    \node [block] (init) {\textbf{A.} \\ Ideal control policy};
    \node [block2, right of=init] (control) {\textbf{B.} \\  QoS-Aware State Estimation };
    \node [block, right of=control] (qos) {\textbf{C.} \\ Control-Aware QoS Adaptation};
    \node [block2, right of=qos] (codesign) {\textbf{D.} \\ Co-design Synthesis };
    \path [line] (init) -- (control);
    \path [line] (control) -- (qos);
    \path [line] (qos) -- (codesign);
\end{tikzpicture}
\caption{Stepwise learning pipeline for wireless edge control co-design problem.}
\label{fig_learning}
\end{figure}
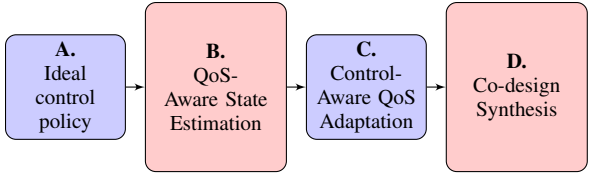

To address the challenges of solving co-design problems under the complexities present in the industrial robotic setting, we propose a modular learning framework that can leverage various machine learning tools to solve individual pieces of the larger co-design problem. Consider that, the co-design policy outputs given by \eqref{eq_q_pol}-\eqref{eq_c_pol} can be broadly separated into (i) the \emph{QoS} policy that informs the management of the wireless network and (ii) the \emph{control} policy governing that directs actuation the plants. We may utilize this natural decomposition present in the problem to separately pre-optimize both components of the co-design policy before proceeding with the joint optimization. In addition to reducing the practical learning complexity, this decomposition allows for various learning or optimization methodologies to be applied throughout the multi-step design process as needed.

The stepwise learning pipeline we utilize for the industrial co-design problem is diagrammed in Fig. \ref{fig_learning}. In each step in the design pipeline, we solve a simpler subproblem using model-free learning algorithms, in particular deep reinforcement learning (DRL). In robotics applications, DRL has been widely used to bypass the difficulties in both modeling the complex dynamics of robotic systems and mathematically formulating the more qualitative nature of task performance. The ability of using model-free learning techniques in each learning step is thus an integral component of co-design of complex edge control systems over wireless networks.

In the following Sections \ref{sec_control}-\ref{sec_e2e}, we detail the decomposition of the co-design problem in \eqref{eq_problem}-\eqref{eq_const} into each of the modules illustrated in Figure \ref{fig_learning}. For each module, we further detail the relevant learning techniques to be applied for the given sub-problem, thus creating and end-to-end learning based co-design solution for wireless industrial systems.

\subsection{Ideal Control Policy}\label{sec_control}

The first module considers the design of an ideal control policy to perform the given control task or satisfy the given performance requirement. That is, we optimize a policy $\hbpi^c_i(\bbx_{i,t}; \hbtheta^c_i)$ that determines the selection of control inputs $\bbu_{i,t}$ given \emph{state information} $\bbx_{i,t}$. Due to the lack of any impact of wireless QoS in this case, we seek to directly minimize the control cost, i.e.
\begin{align}\label{eq_control_ideal}
\hbtheta_i^{c*} := &\argmin_{\hbtheta} \E_{\hbpi^c_{\hbtheta}} \left[ \sum_{t=0}^T  L_i \left( \bbx_{i,t}  \right) \right].
\end{align}
The optimization problem given in \eqref{eq_control_ideal} is naturally simpler than the design of a controller under wireless network effects, and can be considered as a standard closed-loop controller design problem. Such problems can be solved in a model-free manner using standard DRL algorithms \cite{haarnoja2018soft}. We note that some common robotics applications, e.g. AMR navigation, may additionally have well-known model-based control algorithms that can be used as the ideal control policy as well.

\subsection{QoS-Aware State Estimation}\label{sec_estimation}

In the second learning module, we proceed to construct the QoS-aware control policy $\bbpi^c_i$ by utilizing the idealized control policy $\hbpi^{c^*}_i$ given in \eqref{eq_control_ideal} with a QoS-aware \emph{state estimation}. Consider a policy $ \bbpi^e_i(\bby_{i,t}, \delta_{i,t},\tau_{i,t}; \bbtheta^e_i)$ that, for plant $i$, takes as inputs the observation $\bby_{i,t}$ with the QoS performance indicators $\delta_{i,t}$ and $\tau_{i,t}$ and returns an estimate of the current state $\bbx_{i,t}$. State estimation in model-based systems can be performed with, e.g., extended Kalman filters. In the model-free context, however, we may formulate an estimation optimization problem to learn the estimation network parameter $\bbtheta^{e_i}$. 

A natural loss function for estimation is the mean squared error (MSE) loss, with the optimal estimation network parameter subsequently defined as
\begin{align}\label{eq_control_estimation}
\bbtheta_i^{e*} := &\argmin_{\hbtheta} \E_{\hbpi^c_{\hbtheta}, \ccalT} \left[ \sum_{t=0}^T  \| \bbx_{i,t} - \bbpi^e_i(\bby_{i,t}, \zeta_{i,t}; \bbtheta^e_i) \|^2 \right].
\end{align}
In \eqref{eq_control_estimation}, we learn a state estimation policy by employing DRL algorithms under the MSE loss. Note that, to explore the estimation performance under different delays $\zeta_{i,t}$, it is necessary to solve \eqref{eq_control_estimation} under a preselected distribution $\ccalT$ of delays, such as a uniform or Gaussian distribution. 

Composing the estimation policy network with the state-aware control policy $\hbpi^{c^*}_i$, we obtain a complete QoS-aware control policy with parameter $\bbtheta^c_i = [\hbtheta^c_i; \bbtheta^e_i]$, i.e.
\begin{equation}\label{eq_c_pol2}
\bbpi^c_i(\bby_{i,t}, \delta_{i,t},\tau_{i,t}; \bbtheta^c_i) = \hbpi^c_i( \bbpi^e_i(\bby_{i,t}, \delta_{i,t},\tau_{i,t}; \bbtheta^{e*}_i); \hbtheta^{c*}_i) .
\end{equation}
It is worth noting that the decomposition of the optimal QoS-aware control policy into an ideal control policy and estimation policy as presented in \eqref{eq_c_pol2} is not necessarily optimal. Indeed, however, the optimality of this decomposition---known as \emph{certainty equivalence}---can be proven in simple linear systems \cite{bar1974dual, molin2012optimality}. Even this limited theoretical result gives insightful motivation in modularizing the co-design policy in this manner for complex, nonlinear industrial systems.

\subsection{Control-Aware QoS Policy}\label{sec_qos}

With \eqref{eq_control_ideal}-\eqref{eq_c_pol2} we obtain a framework for learning an initial QoS-aware control policy under some distribution of network QoS measures. The next stage in the co-design pipeline is to design a policy wherein we determine appropriate state-aware QoS measures as in \eqref{eq_q_pol}-\eqref{eq_t_pol} that optimize the network efficiency. For this stage, we return to the original policy optimization problem in \eqref{eq_problem}-\eqref{eq_const} but now \emph{restricting our optimization to the dynamic QoS decisions $\pi^q_i(\cdot; \bbtheta^q_i)$ and $\pi^t_i(\cdot; \bbtheta^t_i)$}. The control policy, meanwhile, is fixed as the QoS-aware controller determined in \eqref{eq_c_pol2}. We obtain the following DRL problem to learn the control-aware QoS policy
\begin{align}
\{\bbtheta_i^{q*}, \bbtheta_i^{t*}\} := &\argmin_{\hbtheta} \E_{\hbpi^c_{\hbtheta}} \left \{   \sum_{t=0}^T \left[ C_i \left( \pi_{\bbtheta}(\bby_{i,t})\right) + \right. \right. \nonumber \\
&\qquad \qquad \qquad \left. \left. \lambda L_i \left( \bbx_{i,t}  \right)   -  \lambda J_i^{\max}/T\right] \right\}. \label{eq_q_t_pol2}
\end{align}

Observe in \eqref{eq_q_t_pol2} that we formulate a single loss function from \eqref{eq_problem}-\eqref{eq_const} by placing a cost on constraint violation with some weight $\lambda \geq 0$. As such, the QoS policy parameters $\{\bbtheta_i^{q*}, \bbtheta_i^{t*}\}$ can again be found via standard DRL methods. The optimal QoS policy here seeks the minimum network QoS as a function of the current plant state required to meet the necessary control performance. The weight $\lambda$ that balances low QoS with control performance targets may be tuned heuristically to enforce constraint satisfaction or found algorithmically with primal-dual methods---see, e.g., \cite{lima2020model}.

\subsection{Co-design synthesis} \label{sec_e2e}
The final stage of learning paradigm consists of a \emph{synthesis} of the sequentially learning policies from the prior stages. In particular, the QoS-aware state estimation policy learning in \eqref{eq_control_estimation} is performed prior to the design of the QoS adaptation policy in \eqref{eq_q_t_pol2}. Thus, we may continue the learning of $\bbtheta^e$  in \eqref{eq_control_estimation}, while replacing the generic QoS distribution $\ccalT$ with the QoS adaptation policy specified by $\{\bbtheta_i^{q*}, \bbtheta_i^{t*}\}$. We can likewise repeat the QoS adaptation learning step in \eqref{eq_q_t_pol2} under the new QoS-aware state estimation policy to complete the co-design of these two policies.

\section{QoS-Based Scheduling}\label{sec_scheduling}
In this section we provide a framework for performing wireless scheduling under a set of heterogeneous and dynamic QoS requirements derived via the learned policy in Section \ref{sec_qos}. The scheduling algorithm we develop here contains a number of operations that allow for the centralized wireless scheduling device, called the access point (AP), to minimize resource utilization while meeting the set of QoS metrics of latency $\{\tau_{i,t}\}_{i=1}^m$ and reliability $\{q_{i,t}\}_{i=1}^m$. We emphasize that the approach is generic so as to be applied to various MAC/PHY wireless standards, including modern protocols utilizing OFDMA such as 5G and WiFi 6. As such, we present the scheduling algorithm in terms of a set of high-level mechanisms that do account for the specific options specified in any particular standard. 

Multi-user scheduling under heterogeneous delay and reliability requirements has been well-studied in wireless networks \cite{ergen2003qos, liu2003delay}, including current standards such as 5G \cite{karimi20185g} and WiFi 6 \cite{bankov2017ieee}. To incorporate the highly dynamic and granular values of latency and reliability that are achievable in the co-design framework for industrial robotics, we utilize a two-stage scheduling heuristic capable of addressing any dynamic QoS specification.

Packet delivery reliability is typically controlled via the selection of a modulation and coding scheme (MCS) $\mu$ in modern wireless networks. However, the MCS values utilized in current standards belong to a fixed library $\ccalM$ and are not designed to incorporate highly granular values of packet delivery rate (PDR). To achieve such granularity, we utilize a \emph{probabilistic frame dropping} mechanism \cite{eisen2019control}. That is, to achieve an adaptive PDR value of $q_{i,t}$, we ``drop" the UL or DL frame prior to  entering the queue with probability $\tdq_{i,t}$
\begin{equation}\label{eq_drop}
\tdq_{i,t} = 1 - \frac{q_{i,t}}{q_0},
\end{equation}
where $q_0$ is a fixed link-level PDR achievable under current channel conditions with some MCS $\mu_{i,t} \in \ccalM$. If the frame is not dropped, it is placed in the queue to be transmitted with MCS $\mu_0$.

The second stage in the scheduling method involves a greedy allocation procedure, in which queued packets are transmitted with their associated MCS values in order to achieve the desired delay $\tau_{i,t}$. Specifically, we order the users in order of delivery deadline, and allocate the minimum necessary bandwidth needed under current channel conditions and with MCS to achieve the desired latency. This procedure is intended to maximize users that can  meet deadlines with a standard greedy allocation procedure. Overall, the high level QoS-aware scheduling heuristic can be summarized as 
\begin{enumerate}
    \item For each plant/sensor, drop frame with probability  $\tdq_{i,t}$ for current $q_{i,t}$, otherwise place in queue.
    \item For frames in queue, arrange by delivery deadlines $\tau_{i,t}$ and greedily allocate minimum bandwidth to meet deadline under MCS $\mu_{i,t}$.
\end{enumerate}
The above procedure is generic and can be easily implemented across current wireless standards without significant change to OFDMA scheduling algorithms. In the following section we proceed to demonstrate the performance of learning-based co-design method for an industrial robotics use-case, including simulation in a WiFi 6 network.

\section{Industrial Robotics Simulation}\label{sec_num}

\begin{figure}
\centering
\includegraphics[width=.3\textwidth]{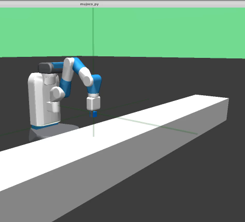} \\
\caption{Conveyor belt and gripping robot in Mujoco environment.}
\label{fig_mujoco}
\end{figure}

\begin{figure}
\centering
\includegraphics[width=.4\textwidth]{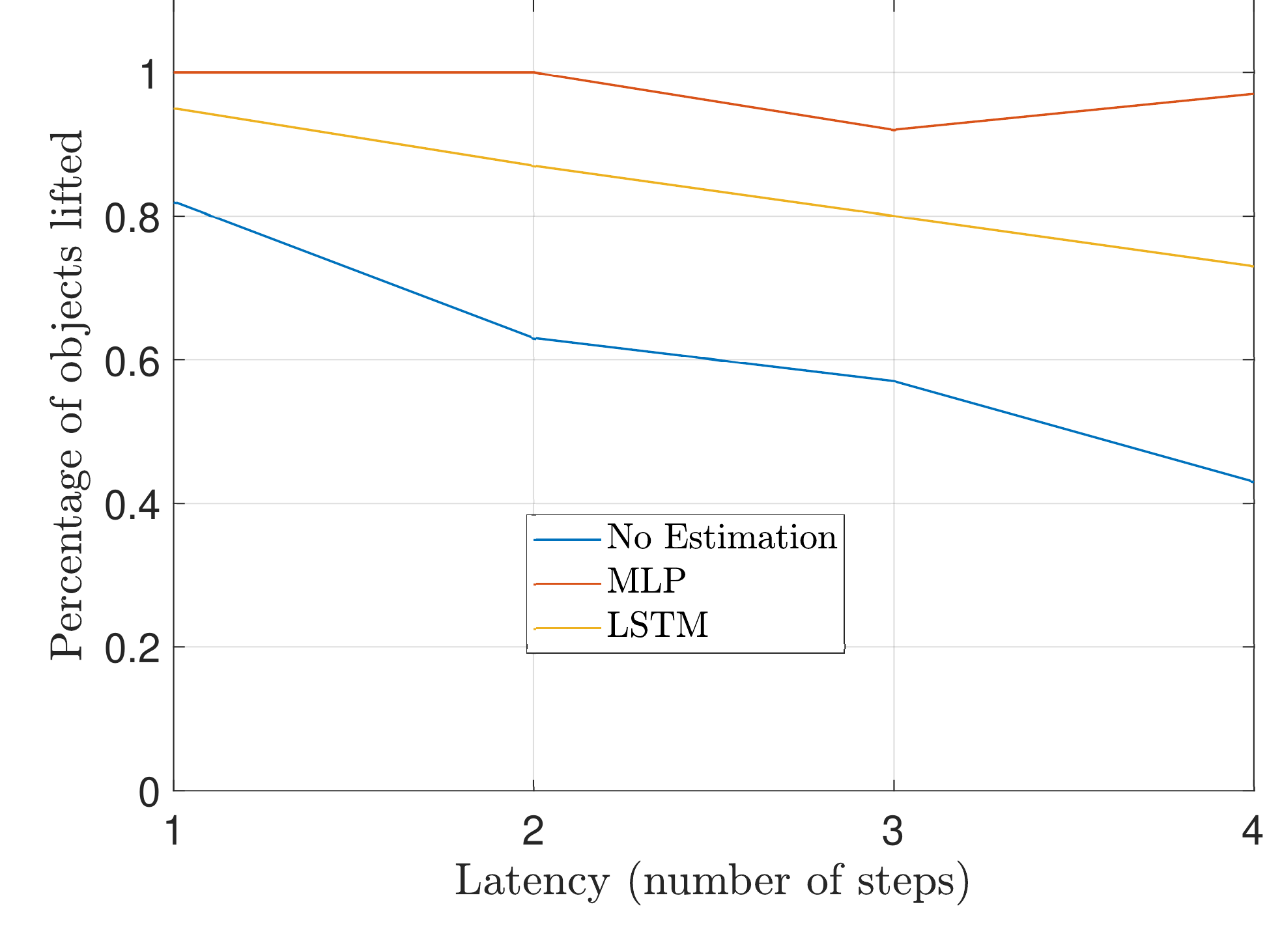} 
\includegraphics[width=.4\textwidth]{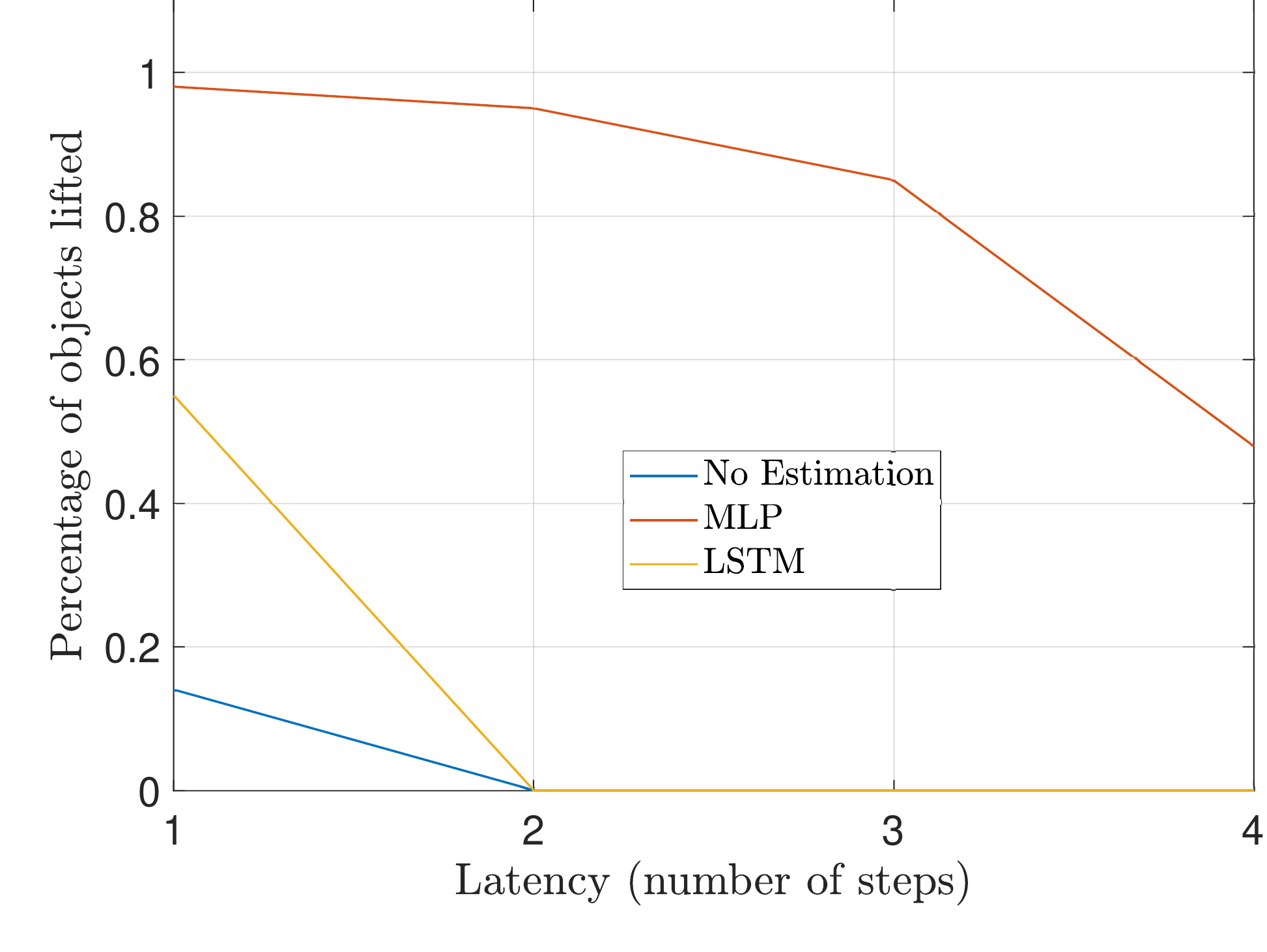}
\caption{Conveyor belt performance under trained QoS-aware state estimation networks. For (top) a 0.2 m/s belt speed, the MLP estimator is able to retain near-optimal task performance. For (bottom) a 0.6 m/s belt speed, the MLP estimator strongly mitigates the performance degradation from high latency.}
\label{fig_cn}
\end{figure}

\begin{figure}
\centering
\includegraphics[width=.4\textwidth]{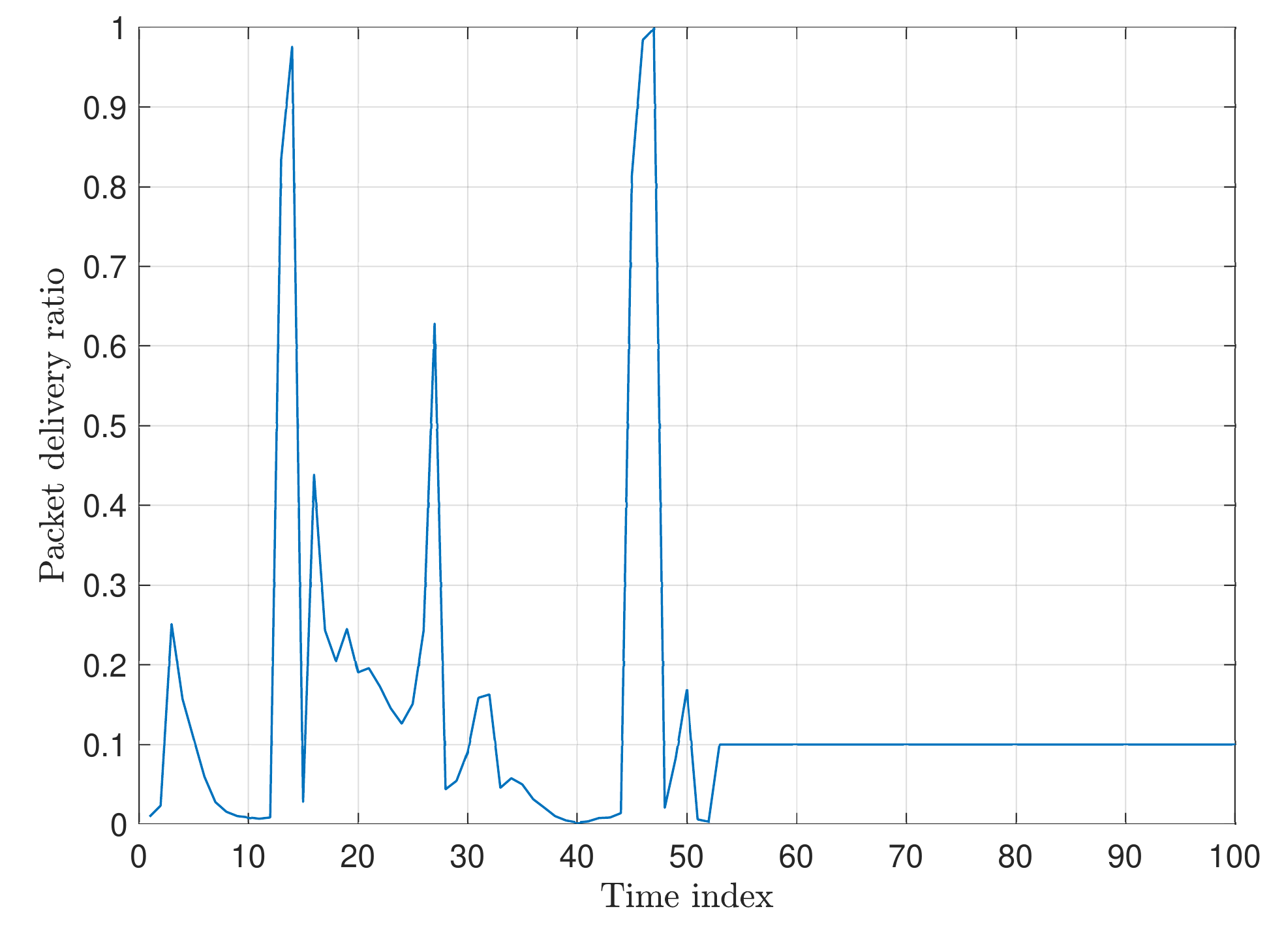}
\caption{Learned QoS (packet delivery ratio) adaptation over episode of Pick and Place task with QoS-aware state estimation. In conjunction with the state estimation policy, the QoS policy dictates that the robot can endure frequent packet drops an successfully lift the object. At time $t=54$, the gripper lifts the object off of the belt switches to a fixed low reliability requirement for the remainder of the episode.}
\label{fig_lat}
\end{figure} 
We evaluate the performance of the proposed QoS-aware co-design methodology in an industrial robotics use case. In particular, we simulate the automated control of a \emph{Pick and Place} task, in which a robotic arm must lift objects off of a moving conveyor belt with an episode length of $T=100$. In this setting, the position of the object is determined from a perception algorithm running at the edge from camera frame inputs, and then transmitted back to the robot for control. The dynamics of such a task can be simulated in the Mujoco simulation environment \cite{todorov2012mujoco}. In Fig. \ref{fig_mujoco}, we show the graphical interface of the gripping robot used in this task. For these simulation results, we consider only the dynamic adaptation of the reliability policy $\pi_i^q(\bby_{i,t-1}, \zeta_{i,t-1}; \bbtheta^q_i)$ while fixing the latency for all users to be a single time step, i.e. $\pi_i^\tau(\bby_{i,t-1}, \zeta_{i,t-1}; \bbtheta^\tau_i) = 1$. Note that, in this numerical example, each time step $t$ corresponds to 40 ms of time. 

As discussed in Section \ref{sec_pf}, we utilize control performance and network resource cost functions in the co-design formulation for a given use-case. In the Pick and Place task, these costs are given by
\begin{align}\label{eq_robot}
C_i(q) &= q, \\
J_i(\bbx) &= \mathbb{I}[\|\bbx^{obj}_{i,t} - \bbx^{grip}_{i,t}\| \leq \eps],
\end{align}
where $\bbx^{obj}_{i,t} - \bbx^{grip}_{i,t}$ is the difference between the object and robotic gripper positions. The control cost $J_i(\bbx)$ is meant to encode the requirement that the robot can successfully lift the object. With these cost functions, we first train the ideal control policy for lifting objects as described in Section \ref{sec_control}. The ideal control policy is trained using the state of the art Soft Actor-Critic (SAC) algorithm \cite{haarnoja2018soft} using the RL Stable Baselines software \cite{stable-baselines}. 

We proceed to train the QoS-aware state estimation policy under a uniform delay distribution $\ccalT$, as described in Section \ref{sec_estimation}. In Figure \ref{fig_cn}, we demonstrate the task performance results of two different state estimation policies in comparison to the performance of the ideal control policy without any state estimation. The first state estimation policy is trained using a multi-layer perception DNN (MLP) and the second policy is trained with an LSTM recurrent neural network policy. In the top figure, we demonstrate for a 0.2 m/s belt speed, that the MLP estimator is able to retain near optimal task performance for up to 4 steps of delay. Likewise, for a faster belt speed of 0.6 m/s, we observe in the bottom figure that the faster belt speed makes the impact of delay on the robotic system much more signifigant. However, the MLP estimator is nonetheless able to mitigate much of the performance degradation as the delay increases. 

Using the dominant MLP estimator, we proceed to train the adaptive QoS policy as described in Section \ref{sec_qos} using the SAC algorithm in Stable Baselines. In Figure \ref{fig_lat}, we demonstrate the output of the QoS policy for a representative successful episode in which the object is lifted from the belt. As can be seen, the reliability, or packet delivery ratio (PDR), selected by the adaptation policy is kept low for successive iterations of the task, while high reliability is needed after many packets are dropped. In the second half of the episode, the robot gripper has grasped the object and the adaptive QoS policy provides fixed low reliability. Moreover, at time $t=54$, the gripper lifts the object off of the belt switches to a fixed low reliability requirement for the remainder of the episode. Overall, Fig. \ref{fig_lat} demonstrates that, while using the QoS-aware estimation policy, the task can be completed for an average of only 13\% packet delivery ratio.

\begin{table}[]
\begin{tabular}{l | ll}
\textbf{Policy}                                        & \textbf{Success Rate} & \textbf{Average PDR} \\\hline
\footnotesize{MLP State est., static QoS}    & 100\%                      & 100\%                \\
\footnotesize{No state est., dynamic QoS} & 78\%                       & 52\%                 \\
\footnotesize{\textbf{MLP State est., dynamic QoS}}    & 98\%                       & 13\%                
\end{tabular}
\caption{ Comparison of performance in Pick and Place task for different learned policies over 100 independent trials. The policies include (i) State Estimation, fixed $100\%$ reliability, (ii) Dynamic reliability Adaptation, no state estimation, and (iii) Dynamic reliability Adaptation and State Estimation (i.e. \textbf{co-design}). In the final case, the complete policy incorporating both QoS-aware state estimation and control-aware QoS adaptation is able to obtain comparable task performance to the baseline with an average of only 13\% PDR.}
\label{tab_p}
\end{table}

We complete the co-design by performing a final synthesis step as described in Section \ref{sec_e2e}. Using the final policy, we perform 100 independent trials and evaluate the performance of the task under different policies. The results are provided in Table \ref{tab_p}. In first row, we report the performance the performance and network efficiency for a state estimation policy without any reliability adaptation. We see, as expected, task performance is perfect, but also having to utilize $100\%$ reliability at all times. In the second row, we demonstrate the performance of a policy in which the reliability is adapted, but without any QoS-aware state estimation. We see a decrease in performance, but allowing for roughly half of the network packets to be dropped. In the final row, we report the performance of our final co-design solution that uses both the QoS-aware state estimation and the state-aware QoS adaptation. Here, we are able to obtain close to ideal task performance, all while using an average of only 13\% packet delivery ratio. These results demonstrate the effectiveness of the co-design and model-free learning methodology in a practical robotics task with high network efficiency, which is key for practical deployment of edge-based wireless control systems.

\subsection{WiFi 6 Network Performance}\label{sec_wifi}
We proceed to use the learned co-design policy to improve the performance of the robotic conveyor belt task in a simulated WiFi 6 network. WiFi 6 features new features for centralized traffic management, such as centrally managed uplink transmissions and OFDMA scheduling---i.e. multiple users transmit simultaneously in orthogonal time/frequency slots \cite{khorov2018tutorial}. In particular, we leverage the QoS-aware greedy scheduling method presented in Section \ref{sec_scheduling} to schedule traffic in a WiFi 6 network that is simulated in Python. The performance of the complete co-design policy is compared against a policy that utilizes QoS-aware state estimation but \emph{no control-aware QoS adaptation}. 

\begin{table}[]
\begin{tabular}{l | l}
Network Bandwidth & 40 MHz \\
RGB Camera frame size (UL) & 13500 bytes \\
Plant state frame size (DL) & 32 bytes \\
Frame arrival period (UL/DL) & 40 ms \\
Schedule update frequency & 1 ms
\end{tabular}
\caption{ WiFi 6 network simulation parameters for conveyor belt system.}
\label{tab_network}
\end{table}

\begin{figure}
\centering
\includegraphics[width=.4\textwidth]{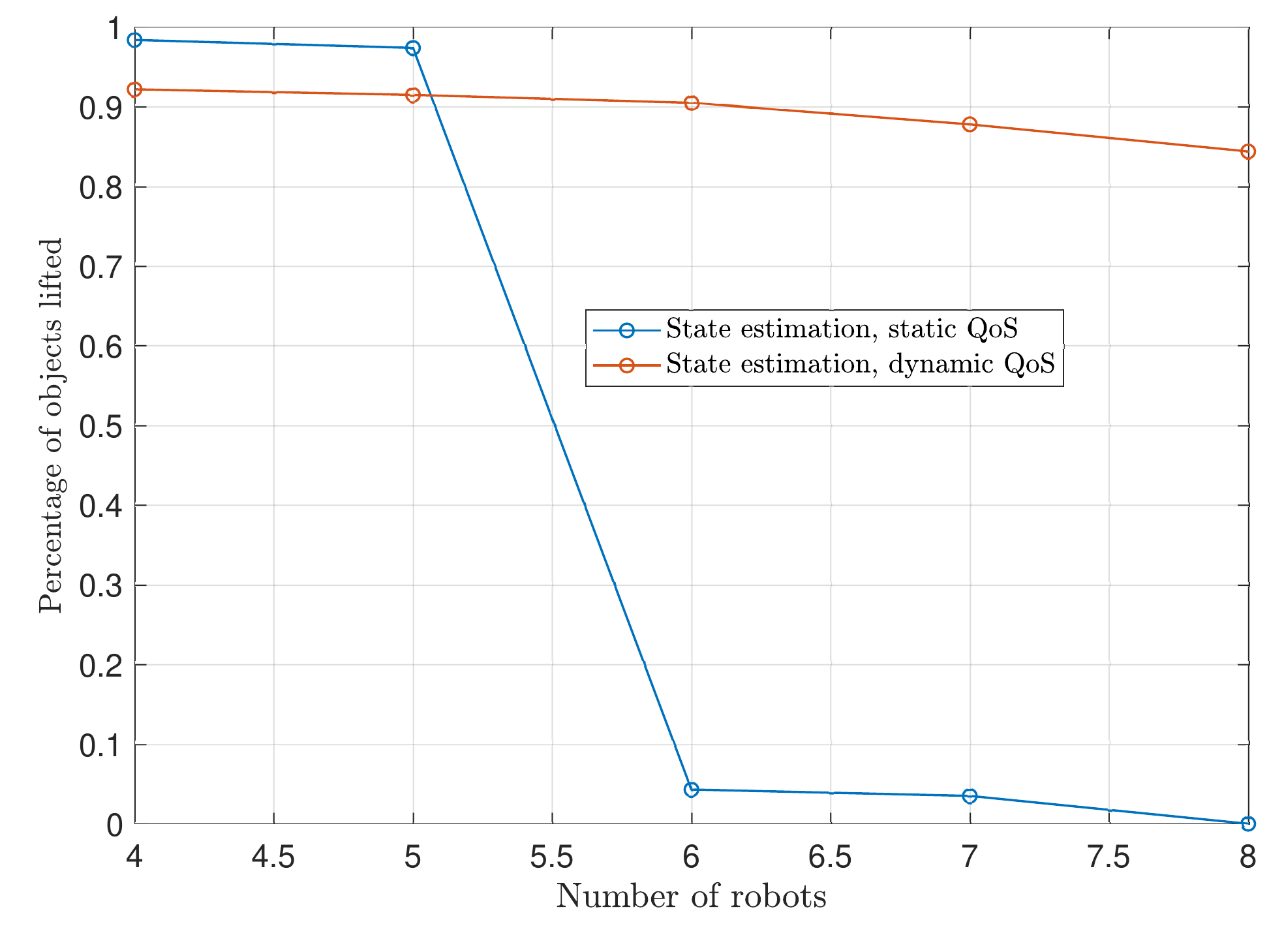} 
\caption{Task Performance in WiFi 6 network using QoS-aware scheduling with and without control-aware QoS adaptation. Without QoS adaptation, the network capacity is exceeded with 6 robotic conveyor belt tasks, while the control-aware dynamic QoS adaptation is able to severely mitigate the performance loss by increasing network capacity.}
\label{fig_wifi}
\end{figure}

As illustrated in Figure \ref{fig_wcs}, the system features a sensor, i.e. camera, sending frames in an uplink (UL) transmission to an Access Point (AP) connected to the edge processor. The state information is processed from the camera frames and sent to the robotic plant in a downlink (DL) transmission. The specifics of the WiFi network environment and traffic details for the conveyor belt system are reported in Table \ref{tab_network}. In Figure \ref{fig_wifi} we report the task performance in a fixed 40 MHz WiFi 6 network under both the complete co-design solution (State estimation and dynamic QoS adaptation) and estimation-only policies. We observe that, without dynamic QoS adaptation, the network traffic exceeds the capacity of the channel, causing significant packet loss and task failure. Using the QoS-aware state estimation in conjunction with the control-aware QoS adaptation, however, allows for the effective capacity of the network the increase and the successful operation of more robotic conveyor belt tasks. These results align with the QoS adaptation policy illustrated in Figure \ref{fig_lat}, which demonstrates the the conveyor belt can be controlled with low QoS when adapting to the state of the object and gripper.

\section{Conclusion}

In this paper we develop a communications-control co-design methodology for dynamically adapting the QoS and control inputs provided to various wireless control loops based on the state and task of the system. The resulting co-design optimization is advantageous because it is inherently scalable and agnostic to the underlying communication network, making it applicable across a wide range of modern wireless protocols and edge deployments. The optimization is practically challenging, however, for modern control systems such as robotics. We further develop a model-free learning framework that allows us to solve individual policies, e.g. ideal control, QoS-aware state estimation, and adaptive QoS, that when put together compose the overall co-design policy. Our numerical results in a industrial conveyor belt task demonstrates the resource gains and performance benefits of our co-design solution.

\urlstyle{same}
\bibliographystyle{IEEEtran}
\bibliography{wireless_ll_control,scheduling_control}

\end{document}